\documentclass[12pt,preprint]{aastex}

\slugcomment{To appear in Ap.\ J., \hspace{5mm} v. \today}

\shorttitle{The stellar population of RSGC1}
\shortauthors{Davies et al.}
\usepackage{graphicx}

\usepackage{lscape}
\usepackage[bf,small]{caption}

\newcommand{\microns}{$\mu$m}
\def\EW{$W_{\lambda}$}
\def\brg{Br\,$\gamma$}
\def\hei{He\,{\sc i}}
\def\feii{Fe\,{\sc ii}}
\def\ga{\mathrel{\hbox{\rlap{\hbox{\lower4pt\hbox{$\sim$}}}\hbox{$>$}}}}
\def\la{\mathrel{\hbox{\rlap{\hbox{\lower4pt\hbox{$\sim$}}}\hbox{$<$}}}}

\def\msun{$M$\mbox{$_{\normalsize\odot}$}}

\def\lsun{$L$\mbox{$_{\normalsize\odot}$}}
\def\kms{\,km~s$^{-1}$}
\def\EW{$W_{\lambda}$}
\def\arcsec{$^{\prime \prime}$}
\def\arcmin{$^{\prime}$}

\begin{document}
\title{The cool supergiant population of the massive young \\ star
cluster RSGC1} \author{Ben Davies\altaffilmark{1}, Don
  F.\ Figer\altaffilmark{1}, Casey J.\ Law\altaffilmark{2}, Rolf-Peter
Kudritzki\altaffilmark{3}, Francisco Najarro\altaffilmark{4},
Artemio Herrero\altaffilmark{5} and John W.\ MacKenty\altaffilmark{6} }


\affil{$^{1}$Chester F.\ Carlson Center for Imaging Science, Rochester
Institute of Technology, 54 Lomb Memorial Drive, Rochester NY, 14623,
USA} 

\affil{$^{2}$Astronomical Institute "Anton Pannekoek", University of
  Amsterdam, Kruislaan 403, 1098 SJ, Amseterdam, The Netherlands}

\affil{$^{3}$Institute for Astronomy, University of Hawaii, 2680
Woodlawn Drive, Honolulu, HI, 96822, USA} 

\affil{$^{4}$Instituto de Estructura de la Materia, Consejo Superior
  de Investigaciones Cientificas, Calle Serrano 121, 28006 Madrid,
  Spain.} 

\affil{$^{5}$Instituto de Astrofísica de Canarias, Via L\`{a}ctea S/N,
  E-38200 La Laguna, Tenerife, Spain} 

\affil{$^{6}$Space Telescope Science Institute, 3700 San Martin
Drive, Baltimore, MD 21218}

\begin{abstract}
We present new high-resolution near-IR spectroscopy and OH maser
observations to investigate the population of cool luminous stars of
the young massive Galactic cluster RSGC1. Using the 2.293\micron\
CO-bandhead feature, we make high-precision radial velocity
measurements of 16 of the 17 candidate Red Supergiants (RSGs)
identified by Figer et al. We show that F16 and F17 are foreground
stars, while we confirm that the rest are indeed physically-associated
RSGs. We determine that Star F15, also associated with the cluster, is
a Yellow Hypergiant based on its luminosity and spectroscopic
similarity to $\rho$~Cas. Using the cluster's radial velocity, we have
derived the kinematic distance to the cluster and revisited the stars'
temperatures and luminosities. We find a larger spread of luminosities
than in the discovery paper, consistent with a cluster age 30\% older
than previously thought (12$\pm$2Myr), and a total initial mass of
$(3\pm1) \times 10^{4}$\msun. The spatial coincidence of the OH maser
with F13, combined with similar radial velocities, is compelling
evidence that the two are related. Combining our results with recent
SiO and H$_2$O maser observations, we find that those stars with maser
emission are the most luminous in the cluster.  From this we suggest
that the maser-active phase is associated with the end of the RSG
stage, when the luminosity-mass ratios are at their highest.
\end{abstract}

\keywords{open clusters \& associations, supergiants, stars:evolution,
stars:late-type, masers}


\section{Introduction} \label{sec:intro}
The Red Supergiants (RSGs) represent a key evolutionary phase in the
life-cycle of stars with initial masses of $\sim$8--30\msun\
\citep[e.g.][]{Mey-Mae00}. Though comparatively brief, the
mass-loss rate in this stage can be many orders of magnitude greater
than on the main-sequence \citep[e.g.][]{Repolust04,v-L05}, and the
mass lost in the RSG phase can determine the terminal mass of the
star, the appearance of the supernova (SN) explosion, and nature of
the stellar remnant \citep[e.g.][]{Heger03}.

The study of RSGs is hampered by low-number statistics -- until
recently only $\sim$200 were known in the Galaxy, and around 40 in the
Large Magellanic Cloud (LMC). Further, the majority of these stars are
isolated, and hence have a variety of ages, initial masses and
metallicities. Ideally, we would like to study large numbers of RSGs
{\it coeval clusters}, where we can be confident that the variables of
metallicity and initial stellar mass are fixed.

Two recent discoveries now present us with the opportunity to study
statistically-significant numbers of RSGs in such clusters. In
\citet[][hereafter FMR06]{Figer06} the discovery of an apparent
cluster of 14 RSGs was presented; while \citet[][hereafter
DFK07]{RSGC2paper} report on the remarkable stellar population of a
second cluster in the same region, which it was shown contains 26
RSGs. In order to use these two clusters as testbeds with which to
study the pre-SN evolution of massive stars, we must first
quantitatively study the properties of the stars within each.

In the case of the second cluster, hereafter RSGC2 (also known as
Stephenson~2), DFK07 used high-resolution spectroscopy to obtain
accurate radial velocities of many stars in the region of the cluster,
and were able to separate proper cluster-members from
background/foreground stars. Further, from the radial velocities they
made a quantitative discussion of the kinematic distance to the
cluster, thus enabling them to determine the stars' luminosities,
temperatures, ages and initial masses.

The details of the stellar properties of the first cluster, hereafter
RSGC1, are less well constrained. In the discovery paper FMR06 showed
with low-resolution spectra that the stars were all late-type; but
with no radial velocity information, they relied on the stars' similar
near-IR colours to argue that the stars were all at the same
distance. They determined this distance by associating a nearby OH
maser, detected by \citet{Blommaert94}, with the RSGs. However this
maser was {\it singly-peaked} -- such masers are formed in the winds
of RSGs and hence are typically {\it doubly-peaked}, with separation
twice the outflow speed. If the second peak was missed in the original
observations, this would lead to an incorrect radial velocity for the
cluster, and hence kinematic distance. Consequently, the cluster
membership, as well as the properties of the luminous cool stars of
RSGC1 are poorly-constrained when compared to RSGC2.

Here, we study the cool supergiants of RSGC1 with data of similar
quality to that presented in DFK07 for RSGC2. We present high
resolution spectroscopy of the CO bandhead feature of 16 of the 17
$K$-bright stars in the field identified in FMR06, allowing us to
determine accurate radial velocities of the stars, and establish the
cluster membership. We also present new observations of the OH maser
source OH25.25-0.16, showing that it is indeed doubly-peaked, and that
the central radial velocity of the profile is consistent with the
average stellar radial velocity. We use this value, in conjunction
with contemporary Galactic rotation curve parameters, to reappraise
the distance to the cluster, the stars' temperatures and luminosities,
and the age and intial mass of the cluster.

We begin in Sect.\ \ref{sec:obs} with a description of the
observations and data reduction steps, and we describe the results and
analysis of the data in Sect.\ \ref{sec:res}. In Sect.\ \ref{sec:disc}
we derive the cluster's age and mass, and compare with similar
analyses of RSGC2. Finally, we use the stellar population of RSGC1 to
investigate the maser-active phase of RSGs.


\section{Observations \& data reduction} \label{sec:obs}

\begin{table}[t]
  \caption{Data for the stars observed. Designations are from FMR06,
  coordinates and apparent magnitudes are from 2MASS. }
  \label{tab:obsdata} \centering
  \begin{tabular}{lccccccccc}
\hline \hline
ID & $\alpha$ & $\delta$ & m$_{\rm J}$ & m$_{\rm H}$ & m$_{\rm K_{S}}$ \\
  & \multicolumn{2}{c}{J2000} & & & \\
\hline
F01 & 18 37 56.29 & -6 52 32.2 &  9.748 &  6.587 &  4.962 \\
F02 & 18 37 55.28 & -6 52 48.4 &  9.904 &  6.695 &  5.029 \\
F03 & 18 37 59.72 & -6 53 49.4 &  9.954 &  6.921 &  5.333 \\
F04 & 18 37 50.90 & -6 53 38.2 &  9.658 &  6.803 &  5.342 \\
F05 & 18 37 55.50 & -6 52 12.2 & 10.547 &  7.178 &  5.535 \\
F06 & 18 37 57.45 & -6 53 25.3 &  9.866 &  7.038 &  5.613 \\
F07 & 18 37 54.31 & -6 52 34.7 &  9.941 &  7.065 &  5.631 \\
F08 & 18 37 55.19 & -6 52 10.7 & 10.772 &  7.330 &  5.654 \\
F09 & 18 37 57.77 & -6 52 22.2 & 10.262 &  7.240 &  5.670 \\
F10 & 18 37 59.53 & -6 53 31.9 & 10.179 &  7.218 &  5.709 \\
F11 & 18 37 51.72 & -6 51 49.9 & 10.467 &  7.325 &  5.722 \\
F12 & 18 38  3.30 & -6 52 45.1 & 10.143 &  7.238 &  5.864 \\
F13 & 18 37 58.90 & -6 52 32.1 & 10.907 &  7.716 &  5.957 \\
F14 & 18 37 47.63 & -6 53  2.3 & 10.495 &  7.576 &  6.167 \\
F15 & 18 37 57.78 & -6 52 32.0 & 10.651 &  8.070 &  6.682 \\
F16 & 18 38  1.30 & -6 52 52.0 & 13.617 &  9.608 &  7.558 \\
F17 & 18 37 48.77 & -6 53  7.7 & 12.763 & 10.188 &  9.003 \\
\hline
  \end{tabular}
\end{table}

\subsection{Radio observations \& data reduction}
OH maser observations at 18 cm were carried out with the
VLA\footnote{The Very Large Array (VLA) is operated by the National
Radio Astronomy Observatory under cooperative agreement with the
National Science Foundation.} on 25 May, 2006 in the AB configuration.
The 256-channel spectrum had a bandwidth of 1.56 MHz giving a velocity
resolution of 1.1 km s$^{-1}$ and was centered at $v_{\rm{LSR}}=100$
km s$^{-1}$.  The observation consisted of a single pointing with
J2000 coordinates of (18h37m52s,-06\degr53\arcmin40\arcsec) and had a
field of view of about 27\arcmin.  The beam size is
3.5\arcsec$\times$1.8\arcsec.  The total integration time was 340
minutes, giving an rms error of about 4 mJy beam$^{-1}$ per 4 kHz
channel.

The data were calibrated using the AIPS package.  The flux and
bandpass calibrator was 1331+305, while 1822-096 was used as a phase
calibrator.  The phase calibrator was mildly resolved in this
observation, so baselines with uv distances greater than 50 $k\lambda$
were removed when solving for a phase solution.  Minimal flagging was
required to remove interference.  The astrometric accuracy of our
positions is dominated by the uncertainty in the position of our phase
calibration source, which is 0.3\arcsec.  The compact source
GPSR5 25.266-0.161 has a J2000 position of (18h37m57.9934s,
-06\degr53\arcmin30.973\arcsec), consistent with previous measurements
with an astrometric accuracy of ~2\arcsec\ \citep{Becker94}.

\subsection{High-resolution spectroscopy}
\subsubsection{Observations}
Observations were taken with {\it NIRSPEC}, the cross-dispersed
echelle spectrograph mounted on Keck-II, during the night of 5$^{th}$
May 2006. We observed stars F01 -- F17, with the exception of
F12 which was missed due to time constraints. F15 was
re-observed in a separate observing run on 12$^{th}$ Aug 2006, using
the same technical setup as described below. Table \ref{tab:obsdata}
lists the coordinates and 2MASS magnitudes of the 17 stars, as
identified in FMR06. The stars' locations are illustrated in Fig.\
\ref{fig:ohpic}.

We used the NIRSPEC-7 filter and 0.576\arcsec$\times$24\arcsec\
slit. Setting the dispersion angle to 62.53\degr, and cross-disperser
angle to 35.53\degr, this gave us a spectral resolution of
$\sim$17,000 in the wavelength range 1.9--2.4\microns. We integrated
for 20s in two nodded positions along the slit for each star. We
observed the B0V star HD~171305 as a telluric standard. We used a
continuum lamp to obtain flat-field frames, while for wavelength
calibration, we observed Ar, Ne, Xe and Kr lamps to obtain as many
spectral lines as possible in the narrow wavelength range. To sample
the gaps between these lines, we also observed the continuum lamp
through an etalon filter.

\subsubsection{Data reduction}
To remove sky emission, dark current and bias level, we subtracted nod
pairs of spectra. Fluctuations in pixel-to-pixel sensitivity were
corrected for by dividing through by the normalized flat-field frame. 

The optics of {\it NIRSPEC} produce spectral orders which are warped
in both the spatial and dispersion directions, and before the spectra
can be extracted this warping must be corrected for in a process known
as {\it rectification}. The methodology we use is the same as that in
DFK07, and is explained in more detail in \citet{Figer03}.  Here we
give a brief summary.

Spatial rectification is done by adding nod-pair spectra and fitting
the spectral traces with a polynomial. Rectification in the dispersion
direction is more complicated; it involves obtaining accurate
wavelengths of the etalon lines using the arc frames, and assuming
that the wavelength of the $n$th order etalon-line $\lambda_{n}$ is
governed by the separation of the etalon plates $t$ via the relation
$\lambda_{n} = t/2n$ (the {\it etalon equation}).

The arc lines were used to get initial estimates of the etalon-line
wavelengths, and hence of the etalon plate separation. The etalon-line
wavelengths were then recomputed using the etalon equation, and used
to re-estimate the wavelengths of the arc lines. The etalon thickness
was fine-tuned in an iterative process until the residuals between the
measured and predicted arc-line wavelengths across all orders were
minimized.

After rectification, the spectra were extraced from each frame by summing
the pixels across the trace in each channel. Shifts between spectra of
up to 4\kms\ ($\la$1 pixel), caused by the star not being quite in the
centre of the slit, were corrected for by cross-correlating the
atmospheric CO$_{2}$ feature at 2.05\microns\ in each spectrum.  

The accuracy of the final wavelength solution is determined from the
residuals between the observed and predicted arc-line wavelengths in
the etalon-fitting process described above, and is better than
$\pm$4\kms. The internal error between spectra, from the CO$_{2}$
telluric feature, is $\ll$1\kms, and so is dominated by systematics in
our analysis process which we estimate to be $\pm$1\kms\ (DFK07).


\section{Results \& analysis} \label{sec:res}

\begin{figure}[t]
  \centering
  \includegraphics[width=12cm,bb=18 180 544 580,clip]{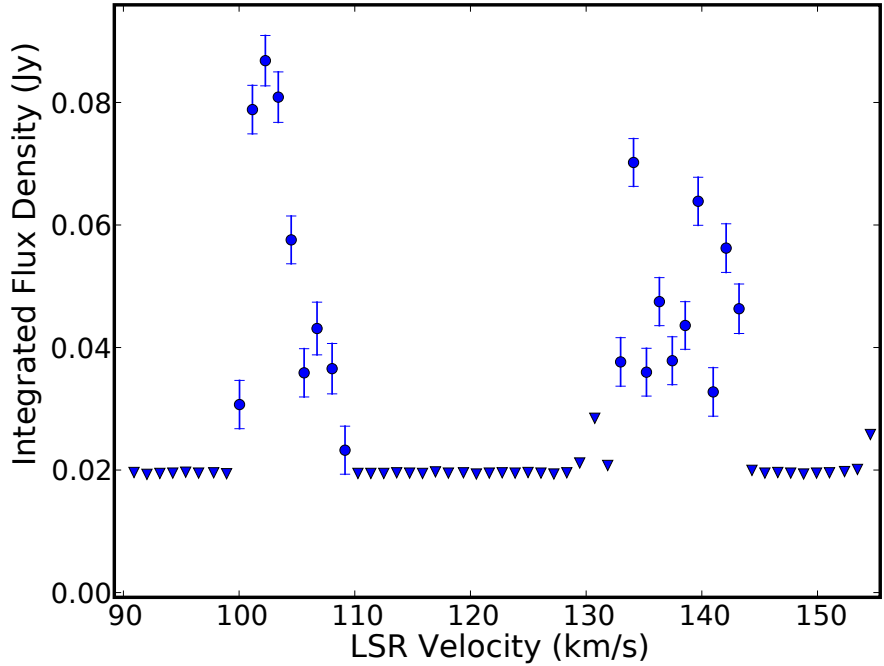}
  \caption{Spectrum at 1612MHz of the source OH25.25-0.16. Our new
  observations clearly show the second peak at
  $\sim$135\kms. Errorbars show $\pm1\sigma$ uncertainty and triangles
  show $5\sigma$ upper limits in the flux density. }
\label{fig:ohspec}
\end{figure}

\begin{figure}[t]
  \centering
  \includegraphics[width=12cm,bb=40 20 850 750,clip]{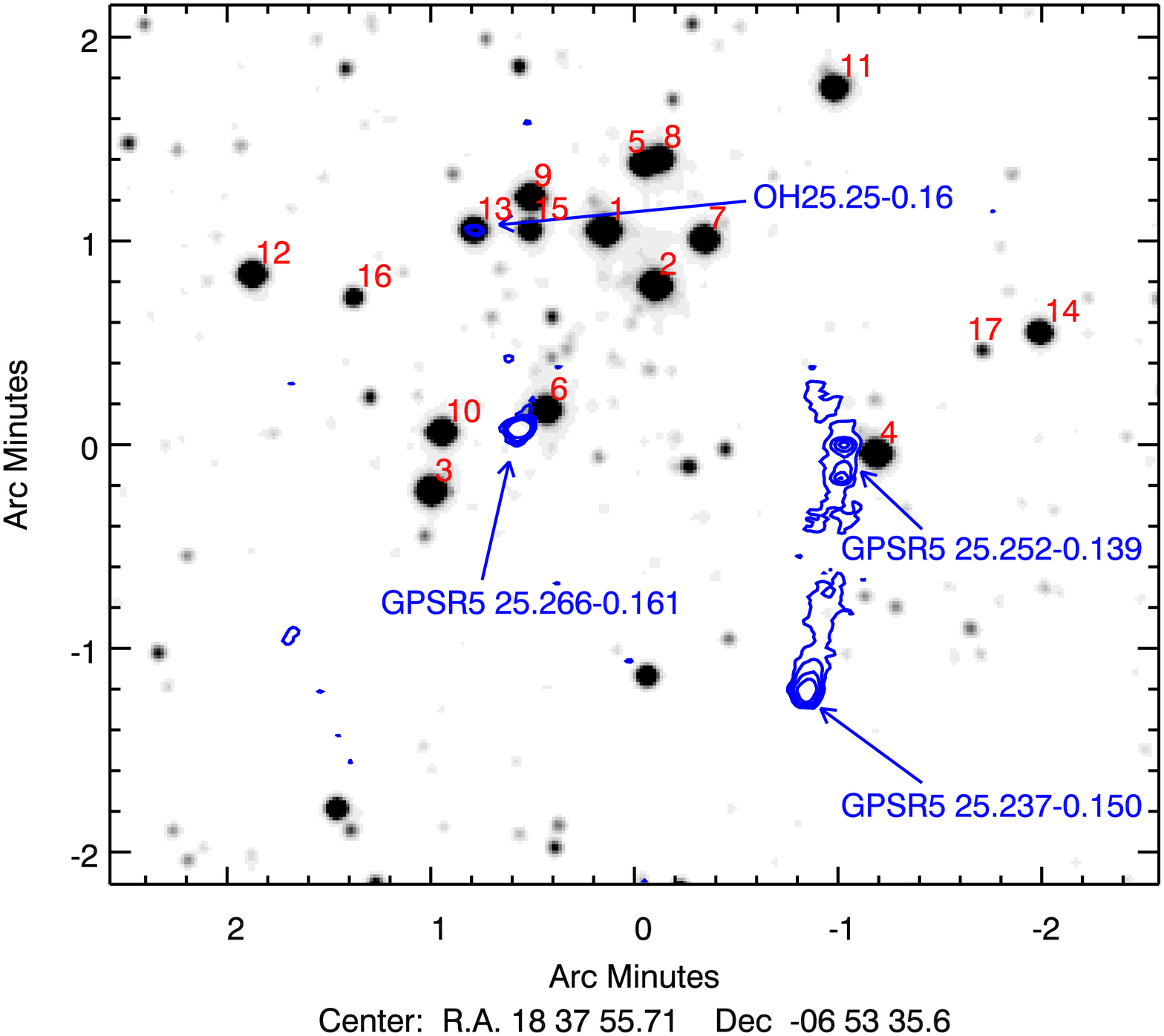}
  \caption{Contour-plot of the 1612Hz radio image, overlayed on
  the 2MASS K$_{S}$-band image of the same region. The image is
  centered on the indicated coordinates (epoch J2000). Contours are
  drawn at 5, 10, 20 and 40$\sigma$ above the background; the
  1$\sigma$ level corresponds to 0.87mJy/beam. The position of the OH
  maser is indicated, and is coincident with star F13. The other
  identified radio sources in the field, labelled in the figure, were
  shown by \citet{T-R06} to be extragalactic in origin. The stars F01
  to F17, as designated by FMR06, are also labelled.}
\label{fig:ohpic}
\end{figure}

\subsection{The maser source OH25.25-0.16} \label{sec:maser}
As mentioned in Sect.\ \ref{sec:intro}, the 1612MHz OH maser forms in
the outflows of RSGs, far above the stellar surface. The velocity
profiles are therefore typically doubly-peaked, with a separation
twice the terminal velocity of the outflow centred on the star's
systemic velocity. However, when observed by \citet{Blommaert94},
OH25.25-0.16 appeared only as a single peak with a radial velocity
$v_{\rm LSR} = 102.2$\kms. 

Our new observation of the OH maser source is shown in velocity-space
in Fig.\ \ref{fig:ohspec}. Here it can be seen that we clearly detect
the second peak. The flux-weighted mean velocities of the two peaks
are $103.8 \pm 0.1$\kms\ and $138.1 \pm 0.9$\kms, calculated using all
channels with emission greater than 5$\sigma$ above the
background. The average velocity of the peaks is $120.9 \pm 0.9$\kms,
consistent with the average velocities of the SiO masers in the
cluster, $120.7 \pm 3.2$\kms, observed by \citet{N-D06}. The implied
outflow speed is 17.1$\pm$0.6\kms, a typical outflow speed for RSGs
\citep{R-Y98}, and similar to the outflow speed of S~Per
\citep[$\sim$16\kms\,][]{Diamond87}, which occupies a similar location
in the HR-diagram as the stars of RSGC1 \citep[][see Sect.\
\ref{sec:disc}]{G-H76}.

In Fig.\ \ref{fig:ohpic} we overlay a contour plot of the 1612MHz
emission on the 2MASS $K_{S}$-band image of the cluster. We find the
positional centroid of the OH maser to be
18$^{h}$37$^{m}$58.882$^{s}$, -06\degr52\arcmin32.28\arcsec\ (J2000),
with a positional uncertainty of 0.3\arcsec. The J2000 position of the
maser is consistent with previous measurement by Blommaert et
al. (1994), which had a positional accuracy of ~4". The maser is also
spatially coincident with star F13, whose 2MASS coordinates are
18$^{h}$37$^{m}$58.908$^{s}$, -06\degr52\arcmin32.11\arcsec\ (J2000),
with positional uncertainty 0.06\arcsec.

The central radial velocity of the OH maser is also consistent with
the SiO observation of F13 by \citet{N-D06}\footnote{This radial
velocity was found from the average velocity of the high- and
low-velocity edges. The centroid of the peak of this source was found
to have a radial velocity of 116.5$\pm$2\kms}, $v_{\rm LSR, F13} =
120.5 \pm 2.0$\kms, and with the CO-bandhead radial velocity
measurement of F13 presented in this paper, 125.4$\pm$4\kms\ (see
Sect.\ \ref{sec:radvs}). From this evidence, it seems highly likely
that the OH maser originates in the outflow of F13.

In addition to the OH maser, we also detect the continuum sources
GPSR5 (25.266-0.161, 25.252-0.139, 25.237-0.15). These sources were
shown conclusively to be extragalactic in origin by \citet{T-R06}.

\begin{figure}[p]
  \centering
  \includegraphics[width=16cm,bb=25 30 560 679,clip]{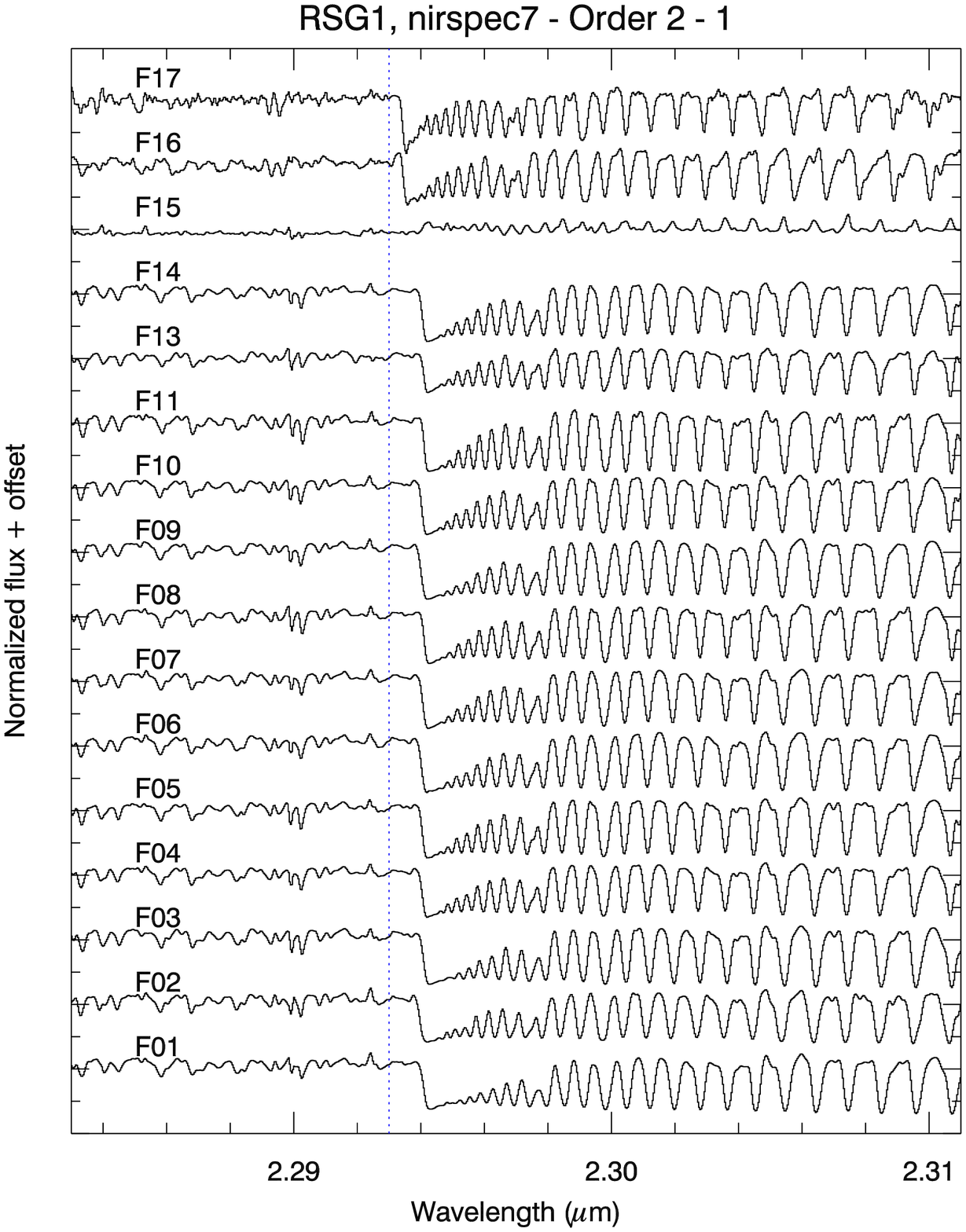}
  \caption{High-resolution spectra of all stars observed in the region
  of the CO-bandhead feature. The dotted blue line indicates the zero
  velocity of the blue edge of the CO bandhead. }
\label{fig:cobandhead}
\end{figure}

\begin{table}[p]
  \begin{center}
    \caption{Observed data for RSGs. (1): the stellar IDs (from
  FMR06); (2): the radial velocity of each star ($\pm$4\kms), and
  the radial velocity measured by \citet{N-D06}, where available
  ($\pm$2\kms); (3): effective temperature; (4): spectral type,
  accurate to $\pm$2 subtypes; (5) derived extinction towards each
  star; (6): absolute $K$-band magnitude; and (7): bolometric
  luminosity. }
  \begin{tabular}{lccccccc}
    \hline \hline
    (1) & \multicolumn{2}{c}{(2)} & (3) & (4) & (5) & (6) & (7) \\
    ID & $V_{\rm LSR}$ & (ND06)$^{1}$ & $T_{\rm eff}$~(K) & Spec Type
    & $A_{K_{S}}$ & $M_{K}$ & log($L_{bol}$/\lsun) \\
     &  \multicolumn{2}{c}{(\kms)} &  &  &  &  &  \\

    \hline
F01 & 129.5 & (117.7) & 3450$\pm$127 & M5 & 2.58$\pm$0.09 & -11.75$^{+0.34}_{-0.30}$ &   5.42$^{+0.12}_{-0.13}$ \\
F02 & 114.2 & (119.7) & 3660$\pm$127 & M2 & 2.83$\pm$0.07 & -11.92$^{+0.34}_{-0.30}$ &   5.56$^{+0.12}_{-0.13}$ \\
F03 & 127.2 & -- & 3450$\pm$127 & M5 & 2.46$\pm$0.09 & -11.28$^{+0.34}_{-0.30}$ &   5.24$^{+0.12}_{-0.13}$ \\
F04 & 121.2 & (124.3) & 3752$\pm$117 & M1 & 2.46$\pm$0.04 & -11.24$^{+0.32}_{-0.28}$ &   5.32$^{+0.11}_{-0.13}$ \\
F05 & 124.8 & -- & 3535$\pm$130 & M4 & 2.77$\pm$0.08 & -11.36$^{+0.34}_{-0.30}$ &   5.29$^{+0.12}_{-0.14}$ \\
F06 & 120.7 & -- & 3450$\pm$127 & M5 & 2.19$\pm$0.09 & -10.70$^{+0.34}_{-0.30}$ &   5.00$^{+0.12}_{-0.13}$ \\
F07 & 121.6 & -- & 3605$\pm$151 & M3 & 2.33$\pm$0.12 & -10.81$^{+0.36}_{-0.32}$ &   5.10$^{+0.13}_{-0.14}$ \\
F08 & 128.2 & -- & 3605$\pm$151 & M3 & 2.84$\pm$0.12 & -11.33$^{+0.36}_{-0.32}$ &   5.30$^{+0.13}_{-0.14}$ \\
F09 & 121.6 & -- & 3399$\pm$150 & M6 & 2.44$\pm$0.08 & -10.92$^{+0.33}_{-0.29}$ &   5.07$^{+0.12}_{-0.13}$ \\
F10 & 122.0 & -- & 3605$\pm$151 & M3 & 2.45$\pm$0.12 & -10.86$^{+0.36}_{-0.32}$ &   5.12$^{+0.13}_{-0.14}$ \\
F11 & 124.1 & -- & 3535$\pm$130 & M4 & 2.63$\pm$0.08 & -11.03$^{+0.34}_{-0.30}$ &   5.16$^{+0.12}_{-0.14}$ \\
F13 & 125.4 & (120.5) & 4015$\pm$140 & K2 & 3.19$\pm$0.09 & -11.39$^{+0.34}_{-0.30}$ &   5.45$^{+0.12}_{-0.13}$ \\
F14 & 122.0 & -- & 3605$\pm$151 & M3 & 2.29$\pm$0.12 & -10.25$^{+0.36}_{-0.32}$ &   4.87$^{+0.13}_{-0.14}$ \\
F15 & 120.8 & -- & 6850$\pm$350 & G0 & 2.65$\pm$0.04 & -10.07$^{+0.40}_{-0.36}$ &   5.36$^{+0.14}_{-0.16}$ \smallskip \\
	{\it F16 } & {\it 42.6 }$^{2}$ & -- & -- & -- & -- & -- & --  \\
	{\it F17 } & {\it 33.2 }$^{2}$ & -- & -- & -- & -- & -- & --  \\
	\hline  \label{tab:rsgparams}
  \end{tabular}
  \end{center}
{$^{1}$\footnotesize We quote the \citet{N-D06} values measured
    by taking the average of the high- and low-velocity edges of the
    maser profiles. \\
$^{2}$Stars F16 \& F17 are determined to be foreground stars, and due  
to the uncertainties in reddening and distance we derive no
    stellar parameters for these stars.}
\end{table}

\subsection{The high-resolution spectra}
The high-resolution observations of the region around the CO-bandhead
feature at 2.293\microns\ for the 17 K-bright stars are shown in Fig.\
\ref{fig:cobandhead}. All stars observed, with the exception of F15,
show the feature strongly in absorption; F15 has it weakly in
emission. Quantitative analysis of the high-resolution spectroscopy
results are described below.

\subsubsection{Stellar radial velocities} \label{sec:radvs}
The radial velocities of each star serve two purposes; firstly they
allow us to distinguish between genuine cluster stars and foreground
stars with similar colours; and secondly they allow us to derive a
kinematic distance to the cluster. 

The differences in radial velocities of the stars can be seen
qualitatively in Fig.\ \ref{fig:cobandhead} -- the blue-edges of the
CO bandhead in stars F16 and F17 are noticably blue-shifted compared
to stars F01 -- F15, which all have very similar radial velocities.

In order to accurately quantify the radial velocities of the stars, we
implemented the same technique as presented in \citet{Figer03} and
DFK07. We cross-correlated the spectra shown in Fig.\
\ref{fig:cobandhead} with that of the high-resolution spectrum of Arcturus presented in \citet{W-H96arct}, which had been degraded to
the same spectral resolution as our data. For star F15, which has CO
in emission, we inverted the spectrum before analysis. We experimented
with isolating different spectral ranges during the cross-correlation,
to determine the robustness of our measurements. We found that the
measured velocities were stable to within $\pm$1\kms, which we take to
be the {\it internal} error between individual measurements. The
absolute uncertainty on the measurements is limited by the accuracy of
the wavelength solution, $\pm$4\kms\ (see Sect.\ \ref{sec:obs}).

We find that stars F01 -- F15 all have radial velocities in the range
$v_{\rm LSR} \sim$115-125\kms, while stars F16 and F17 have $v_{\rm
LSR}$ of 33\kms\ and 43\kms\ respectively. From this, we conclude that
the two faintest stars observed are foreground stars, while stars F01
-- F15 are physical members of the cluster. As F12 was unobserved, the
status of this star is still unclear. We note that the CO absorption
strengths of stars F16 and F17 are much lower than those of the RSGs,
and instead are more typical of less luminous stars. This is
consistent with these stars being foreground objects.

The measured radial velocities of all stars observed are listed in
Table \ref{tab:rsgparams}. Also listed in Table \ref{tab:rsgparams},
where available, are the radial velocities determined from SiO maser
emission by \citet{N-D06}. We see that the two measurements of stars
F02, F04 and F13 are within $\sim$2$\sigma$, while for F01 it is
$\sim$3$\sigma$. As SiO masers are commonly thought to trace the
stellar systemic velocity \citep{Jewell91}, we would expect the two
velocity measurements to agree well. However, we note that the
observations of \citet{N-D06} may have been hampered by their
beam-size; their observations of F01, F02 and F13 have at least one
other RSG within the beam FWHM. In the case of F04, which is well
separated from the other RSGs, the CO bandhead and SiO maser
measurements are in excellent agreement. As mentioned in Sect.\
\ref{sec:maser}, the measured radial velocity of F13 is in excellent
agreement with observations of the OH and SiO maser sources at the
same location.

\subsubsection{Distance to RSGC1} \label{sec:dist}
We take the mean radial velocity of the stars observed at high
spectral resolution and compare with the Galactic rotation curve,
using the contemporary measurements collated by \citet{K-D07}. The
mean radial velocity of the stars from the CO observations is
123.0$\pm$1.0\kms, with an uncertainty determined from Poisson
statistics of the measurements. The absolute uncertainty on the radial
velocity is therefore dominated by that in the wavelength solution,
$\pm$4\kms. This compares well to the average radial velocity found by
\citet{N-D06}, $\sim$120$\pm$2\kms, from their SiO maser
observations\footnote{These values are the average of their two
different methods of measuring the radial velocity of each star from
the maser line profile}, and the central velocity of our new 1612MHz
OH maser observation, $120.9 \pm 0.9$\kms.

\begin{figure}[t]
  \centering
  \includegraphics[width=14cm,bb=47 10 605 480]{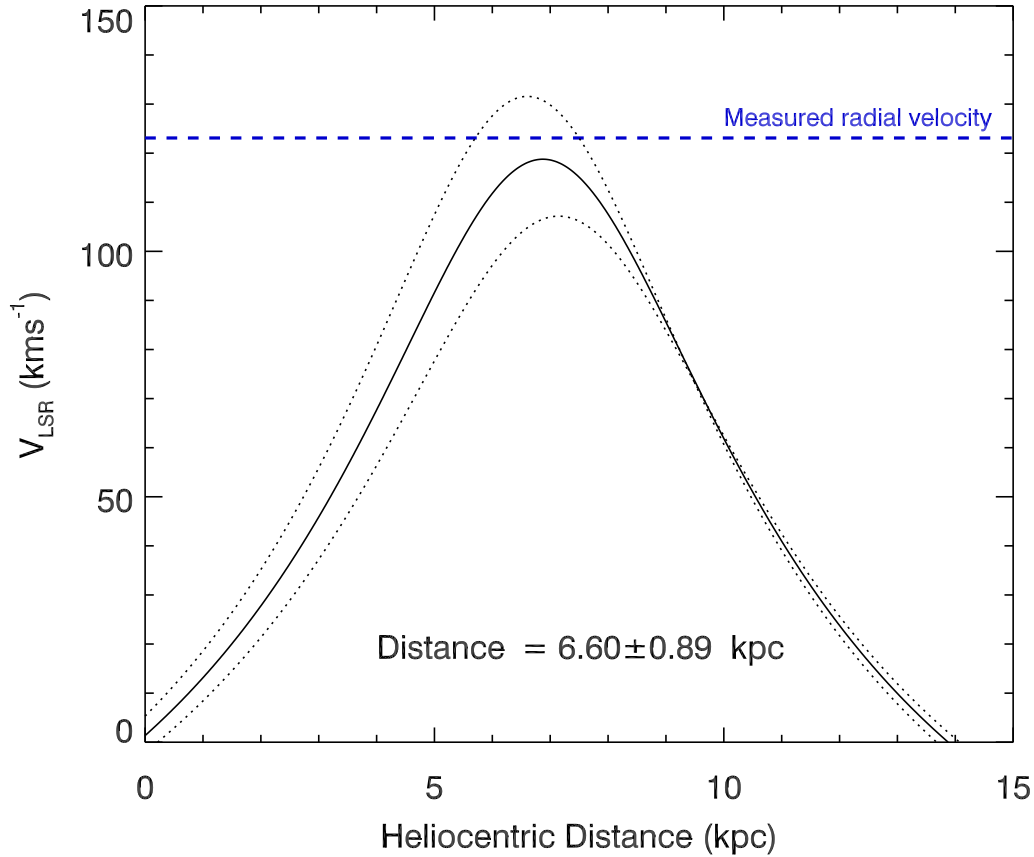}
  \caption{Galactic rotation curve in the direction of RSGC1
  \citep{B-B93}, using the measurements of collated by
  \citet{K-D07}. }
  \label{fig:grotcurve}
\end{figure}

In Fig.\ \ref{fig:grotcurve} we compare our radial velocity to the
Galactic rotation curve in the direction of $l=25.15$\degr,
$b=-0.15$\degr. We use the distance to the Galactic centre $D_{\rm
GAL} = 7.5 \pm 0.3$kpc \citep{Eisenhauer05}, and solar rotational
velocity $\Theta_{\sun} = 214 \pm 7$\kms\ \citep{F-W97,R-B04}. We use
the uncertainties on these values to construct `maximal' and `minimal'
rotation curves in Fig.\ \ref{fig:grotcurve}.

The cluster radial velocity actually extends beyond the
asymptotic point of the curve; however it lies well within the two
`error' curves, and therefore could simply be due to the uncertainties
in $D_{\rm GAL}$ and $\Theta_{\sun}$. To determine the distance to the
cluster, we take the average of the two points where the radial
velocity intercepts the `maximal' rotation curve. This gives us a
kinematic distance to the cluster of 6.60$\pm$0.89\,kpc, slightly
larger than the distance quoted in FMR06 when using the radial
velocity of the singly-peaked OH maser.

\subsubsection{Effective temperatures} \label{sec:temp}
To determine the spectral-types of the RSGs, and hence their effective
temperatures, we used the same empirical method described in
\citet{RSGC2paper}. We compared the equivalent width (\EW) of the CO
bandhead absorption with that of template stars, taken from the
catalogues of \citet{K-H86}, \citet{W-H96} and \citet{W-H97}. We
defined a measurement region of 2.294-2.304\microns, and defined the
continuum as the median of the range 2.288-2.293\microns. We estimated
the uncertainty by repeating the measurements with slightly adjusted
continuum regions, and found that measurements were stable to
$\sim$1\AA, or $\sim$5\%. To find the spectral-types of the RSGs, we
compared the \EW\ measurements with a linear fit to the \EW\ of the
template stars as a function of spectral-type. Using this method, we
are able to determine spectral-types to within $\pm$2 subtypes (see
DFK07). In converting spectral-type to effective temperature, we used
the temperature scale of \citet{Levesque05}. The derived
spectral-types and effective temperatures for the RSGs are listed in
Table \ref{tab:rsgparams}.

For star F15 the method described above breaks down, as this star has
CO in emission. The star's radial velocity suggests that it is part of
the cluster, and hence a supergiant. It was assigned the spectral-type
G6~I in FMR06, and hence deemed to be a `Yellow Hypergiant' (YHG),
based on its weak CO-bandhead absorption. Given that this was only a
marginal detection of CO absorption here we reappraise the star's
spectral type.

\begin{figure}[p]
  \centering
  \includegraphics[width=14cm,bb=25 10 560 720]{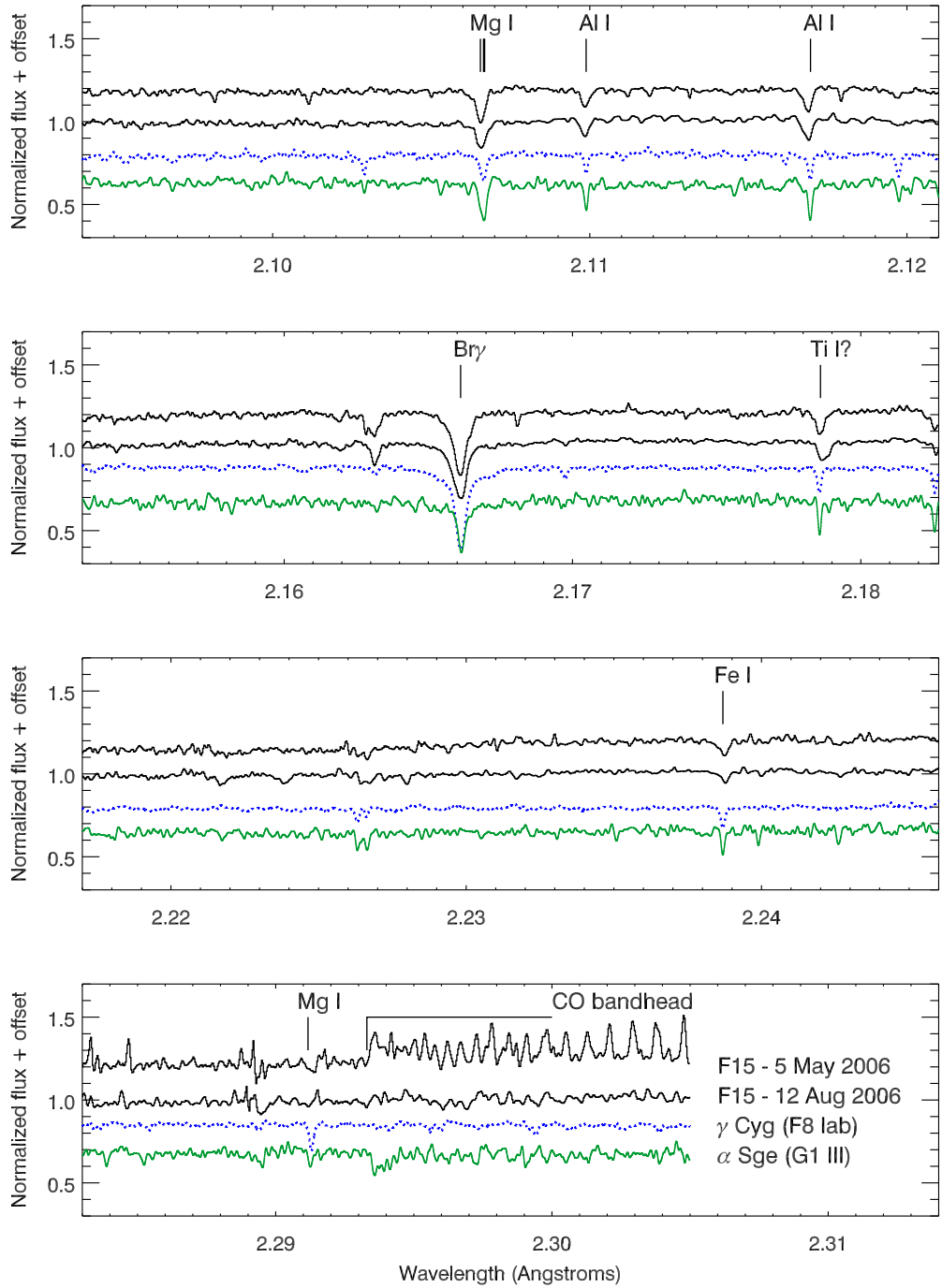}
  \caption{The high-resolution K-band spectrum of F15, compared to the
  spectra of $\gamma$\,Cyg and $\alpha$\,Sge from \citet{W-H97}. }
  \label{fig:yhg}
\end{figure}

Figure \ref{fig:yhg} shows the spectra of F15 on the two occasions it
was observed. It can be seen that the CO-bandhead, which was present
in emission in May, was not detected at all in August. Over the rest
of the star's spectrum, no discernable variability is observed.

For comparison, Fig.\ \ref{fig:yhg} shows similar template spectra of
$\gamma$~Cyg, spectral-type F8~Iab, and $\alpha$~Sge, spectral-type
G1~III, which have been resampled to the same spectral resolution as
our data \citep[taken from][]{W-H97}. Though admittedly $\alpha$~Sge
has a lower luminosity class than that inferred for F15, it serves to
give some insight into F15's temperature. At spectral-type G1, the
CO-bandhead absorption is still seen, albeit weakly. The star also
shows the atomic absorption lines of Al~{\sc i}, Fe~{\sc i}, Mg~{\sc
i}, and \brg. However, at this temperature molecular absorption,
mostly from CN, can be seen as the undulating `noise'-like features
throughout the spectrum. At spectral-type F8, the CN and CO absorption
are gone, while the atomic absorption lines remain.

From these comparison spectra, we assign a spectral-type to F15 of G0,
$\pm$2 subtypes, and hence an effective temperature of $T_{\rm eff} =
6500-7200$K. Indeed, the star is very similar spectroscopically to
$\rho$~Cas, one of the archetypal YHGs, which also shows transient CO
emission / absorption \citep[see spectra presented in][]{Gorlova06}.

\subsubsection{Extinction} \label{sec:extinct}
The extinction towards each star is measured by comparing the 2MASS
infrared colours of the stars to the intrinsic colours of supergiants
with the same spectral-type. For the RSGs we use the observations of
Galactic stars with luminosity class Iab from \citet{Elias85}; while
for F15 (the YHG) we use the tables of \citet{Koorneef83}. We define
the excess between the observed and intrinsic colours as the reddening
towards each star. We convert this reddening to an extinction towards
each star using the relationship of \citet{R-L85},

\begin{equation}
A_{K_{S}} = \frac{E_{\lambda - K_{S}}}
{(\lambda/\lambda_{K_{S}})^{1.53} - 1}
\label{equ:reddening}
\end{equation}

\noindent where $\lambda$ is the wavelength appropriate for the 2MASS
$J$ or $H$ filters. To determine the uncertainty in each extinction
measurement we derive extinctions for the upper and lower limits to
each star's spectral-type. Where the difference between the derived
$A_{K_{S}} (J-K_{S})$ and $A_{K_{S}} (H-K_{S})$ values is outside this
uncertainty, we adopt half this difference as the error in the
measurement.

Using this method, we find a median extinction of $A_{K_{S}} (J-K_{S})
= 2.58$, and $A_{K_{S}} (H-K_{S}) = 2.62$. Each have uncertainties of
0.07mags from Poisson statistics, and are therefore in good agreement
with one another. We adopt the mean of these measurements, $A_{K_{S}}
= 2.60 \pm 0.07$ as the median extinction towards the cluster. We note
that the extinction towards the YHG, $A_{\rm K, F15} = 2.65\pm0.02$,
is consistent with that derived for the RSGs.

The median cluster extinction is slightly lower than the $A_{K_{S}} =
2.74 \pm 0.02$ derived in FMR06. We consider the latest measurement to
be the more reliable, due to the extra assumptions used in FMR06:
instead of dereddening each star according to the intrinsic colours
appropriate for its spectral-type, they dereddened {\it all} stars to
the mean colour of M supergiants from \citet{Elias85}. The ranges in
colours of M supergiants are $\Delta (J-K) \sim$0.3 and $\Delta (H-K)
\sim$0.1, which each correspond to $\Delta A_{K_{S}} \sim$0.2 using
Eqn.\ (\ref{equ:reddening}). Hence, if the spectral-types of the RSGs
are asymmetrically distributed about the mean spectral-type, this may
produce a derived extinction out by as much as 0.2, consistent with
the difference between the extinctions derived here and in FMR06.

\subsubsection{Luminosities} \label{sec:lum}
We take the extinctions toward each star derived in Sect.\
\ref{sec:extinct}, in conjunction with the distance to the cluster
estimated in Sect.\ \ref{sec:dist} to determine the absolute
$K_{S}$-band magnitudes of the stars. To convert these to bolometric
luminosities ($L_{\star}$) we interpolate over the contemporary
bolometric corrections $BC_{K}$ for RSGs, given in \citet{Levesque05},
for the stellar temperatures derived in Sect.\ \ref{sec:temp}.

The absolute uncertainty in each $L_{\star}$ determination is $\sqrt(
\delta A_{K} + \delta BC_{K} + \delta D_{\rm cl})$. As we can
confidently make the approximation that all the stars are all located
at the same distance, the uncertainty in distance $\delta D_{\rm
cl})$ can be neglected when analysing the luminosity spread of the
stars to infer the cluster's age (see Sect.\ \ref{sec:age}). The
uncertainties in both $A_{K}$ and $BC_{K}$ are carried forward from
the error in $T_{\rm eff}$, and are determined by substituting the
upper and lower limits to the stars' temperatures.

We list the stars' luminosities in Table \ref{tab:rsgparams}, along
with uncertainties which include the error in the cluster
distance. The newly-derived values are similar to those quoted in
FMR06, typically within $\pm$0.3dex. As stated above, FMR06 used the
same extinction towards all stars in the cluster, an approximation
which breaks down if there are large variations in the interstellar
extinction across the field or if a star has extra circumstellar
extinction. Also, our high S/N, high-resolution spectra give a more
accurate picture of variations in CO equivalent width, and hence
better constrained stellar temperatures -- key in evaluating the
stars' bolometric corrections. For these reasons, we conclude the
bolometric luminosities derived here to be the more accurate than
those quoted in FMR06.

\subsubsection{Spectral energy distributions}
Using the Galactic plane surveys of MSX and GLIMPSE
\citep{Egan01,Benjamin03}, as well as 2MASS, we have collated IR
photometry for all stars observed here. For the brighter stars, mid-IR
photometry is unavailable due to the stars saturating in the images
(e.g. F01 and F02); while fainter stars in crowded regions (e.g. F15)
are dwarfed by brighter nearby stars (e.g. F09). We rejected all
upper-limit measurements and all detections fainter than 10$\sigma$.

In Fig.\ \ref{fig:seds} we plot the spectral energy distribution (SED)
of each star. Also plotted in the figure are the stars' de-reddened
photometry, which were calculated using the interstellar extinction
toward each star in conjuction with the extinction-laws quoted by
\citet{Indebetouw05} and \citet{Messineo05}. We have overplotted
blackbody curves appropriate for the stellar temperatures and
luminosities calculated in Sects.\ \ref{sec:temp} and \ref{sec:lum}.

In all cases, the blackbodies provide excellent fits to the near-IR
photometry, even in the case of F15 where a less accurate method of
temperature estimation was possible. This serves to validate the
stellar luminosities and temperatures of the stars derived above
(c.f.\ Fig.\ 13 of FMR06, where poorer fits to the IR photometry were
obtained.).

In all cases where photometry is available, the mid-IR MSX data shows
that the stars have considerable excess emission. This is indicative
of warm circumstellar dust, a product of the high mass-loss rates of
the stars. Also, the SEDs appear to show bumps around 12\microns,
which can be understood as silicate emission from the oxygen-rich
dust. A detailed study of the circumstellar material around these
objects will be the subject of a future paper.

\begin{figure}[p]
  \centering
  \includegraphics[width=16cm,bb=10 20 566 566]{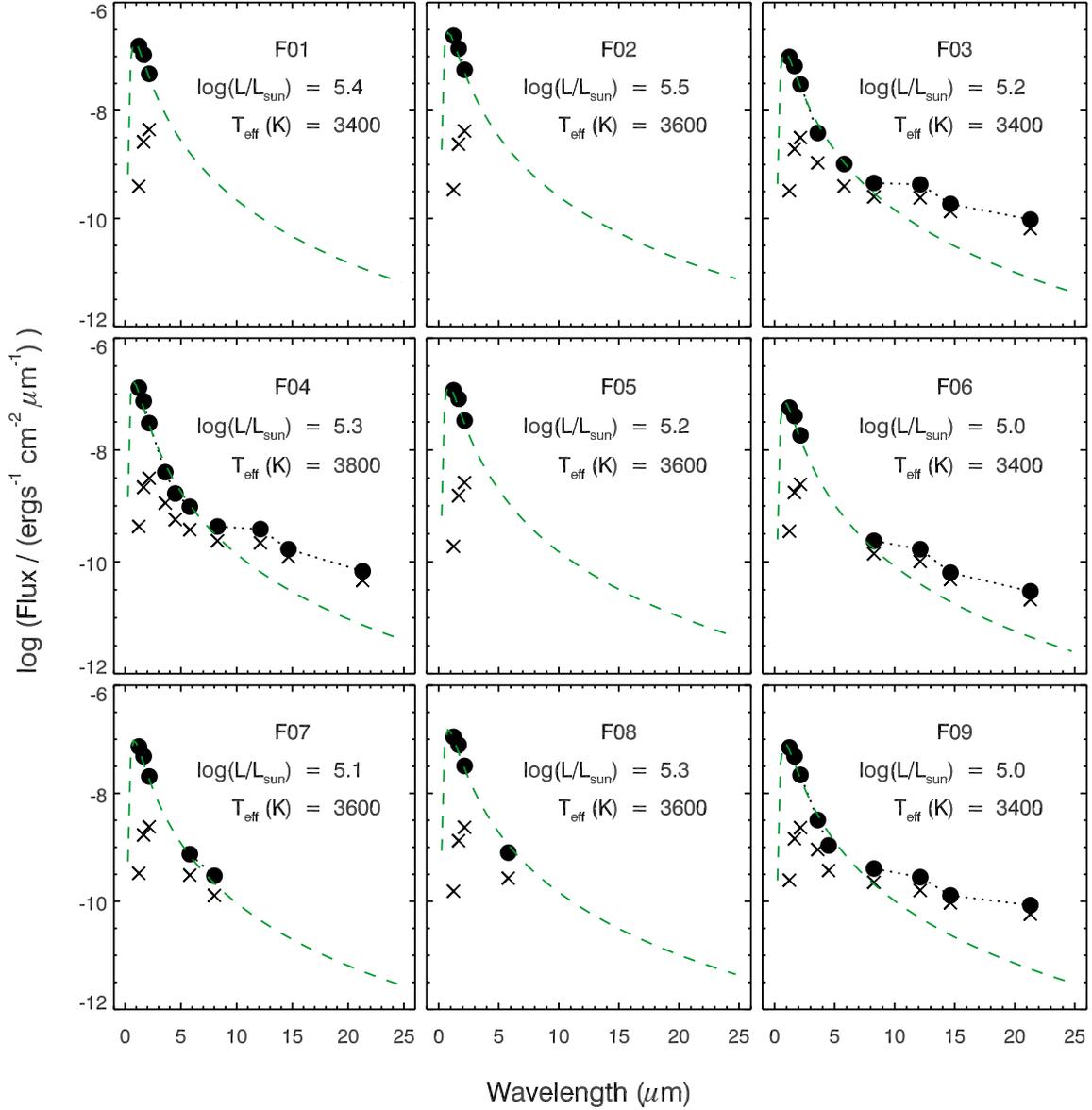}
  \caption{Spectral energy distributions of cluster members. Crosses
  are raw photometry from 2MASS, GLIMPSE \& MSX; filled circles are
  dereddened according to the stellar extictions derived in Sect.\
  \ref{sec:extinct}. Over plotted in green is a black-body curve, with
  $T$ determined from spectral-type of star and luminosity according
  to dereddened K-band magnitude at a distance of 5.88kpc. }
  \label{fig:seds}
\end{figure}
\addtocounter{figure}{-1}
\begin{figure}[t]
  \centering
  \includegraphics[width=16cm,bb=10 180 566 566]{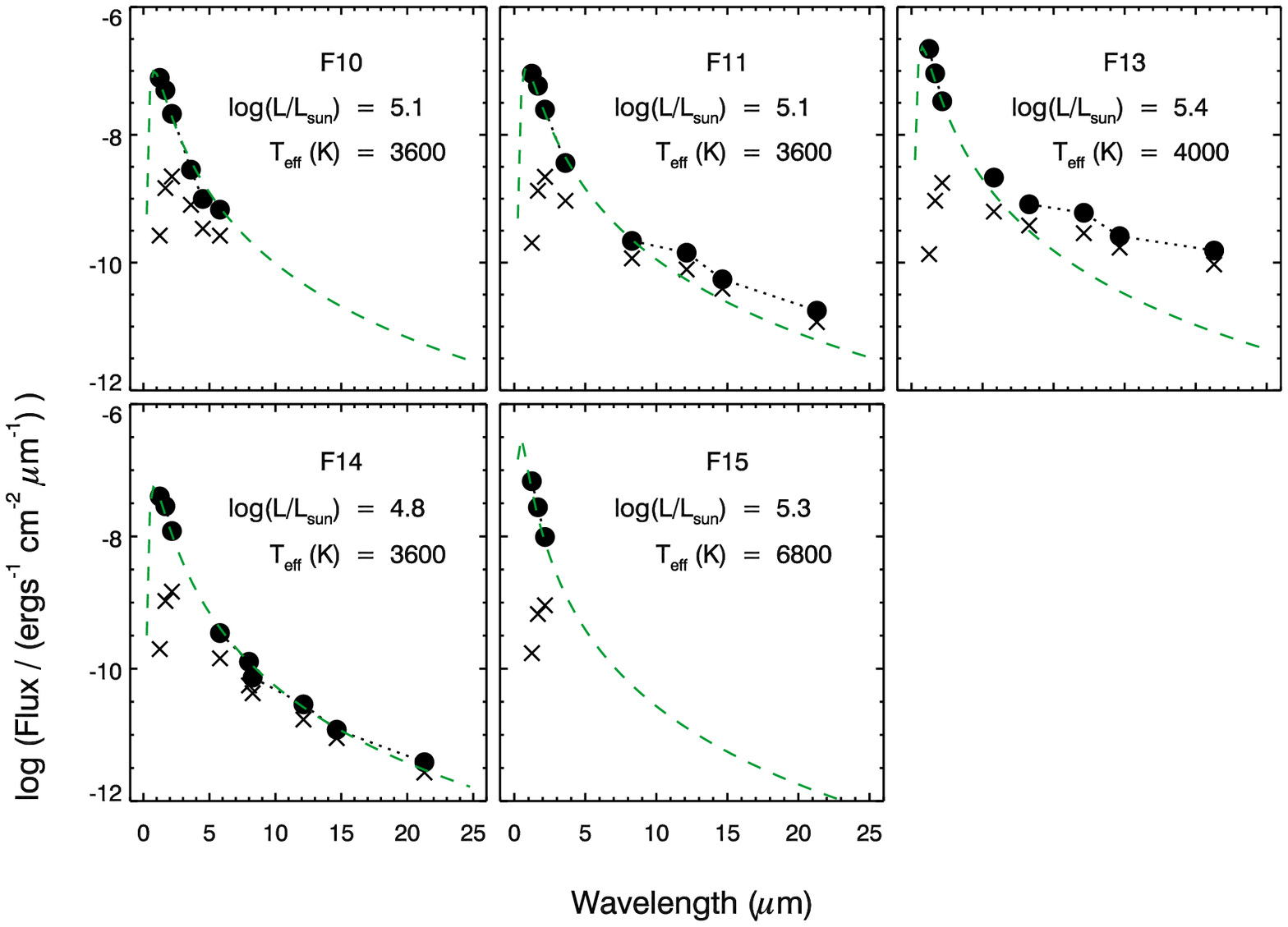}
  \caption{ Cont. }
\end{figure}



\section{Discussion} \label{sec:disc}

\subsection{Cluster age} \label{sec:age}
\begin{figure}[t]
  \centering
  \includegraphics[width=12cm,bb=15 10 560 450]{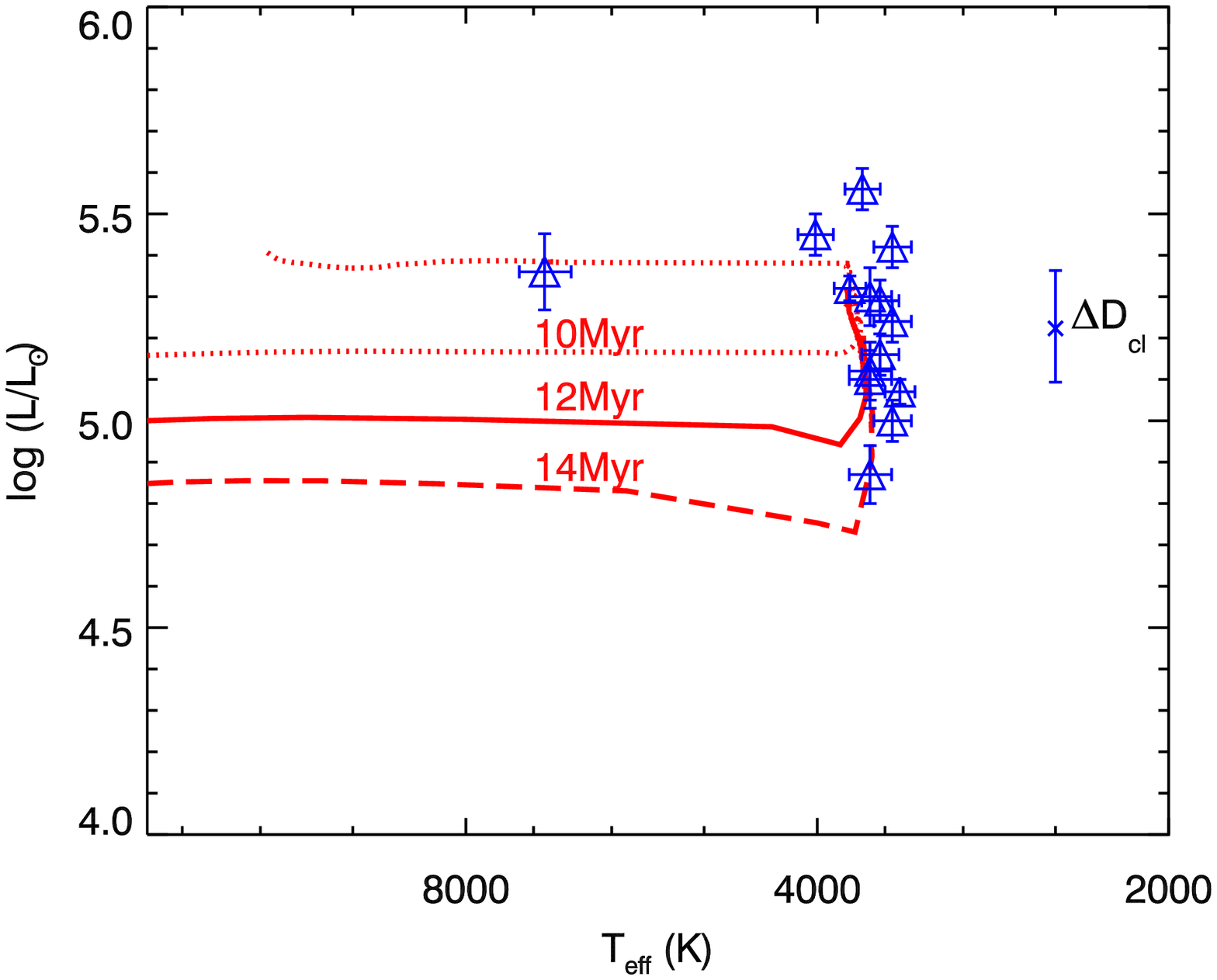}
  \caption{H-R diagram, showing the location of the RSGs and one
  YHG. Overplotted are rotating Geneva isochrones from
  \citet{Mey-Mae00}, with ages 10, 12 and 14 Myr. The luminosity
  spread of the RSGs is well-matched by the 12Myr isochrone. Errors on
  the data-points do not include the uncertainty in cluster distance
  $\Delta D_{\rm cl}$ -- the magnitude of this error is indicated at the
  right of the plot. The data-point separated from the rest is the YHG
  F15. }
  \label{fig:hrd}
\end{figure}

In Fig.\ \ref{fig:hrd}, we plot the derived temperatures and
luminosities of the cool, luminous stars on a H-R diagram. Also
plotted are isochrones taken from the stellar evolutionary models of
\citet{Mey-Mae00} which include the effects of rotation, and have
initial rotational velocity set to 300\kms. As we can confidently make
the approximation that all the stars are at the same distance, we do
not include the error in distance on each data-point. The magnitude of
the error in $L_{\star}$ when this uncertainty {\it is} included is
shown on the right of the plot.

The figure shows that the temperatures and luminosity spread of the
RSGs are well-matched by the 12Myr isochrone. The 10Myr isochrone
cannot reproduce the low-luminosity stars, while the 14Myr isochrone
is too faint to fit the high-luminosity stars. When the uncertainty in
cluster distance is taken into account ($\Delta D_{\rm cl}$ in Fig.\
\ref{fig:hrd}), the 10Myr and 14Myr isochrones appear more
reasonable. We experimented with different evolutionary models, namely
non-rotating models with varying mass-loss rates and metallicity
\citep{Schaller92,Schaerer93,Meynet94}. We found generally that the
non-rotating models gave ages that were $\la$2\,Myr younger. We settle
on an age estimate for RSGC1 of 12$\pm$2Myr.

We note that the YHG F15 does {\it not} lie on the 12Myr
isochrone in Fig.\ \ref{fig:hrd}. Fitting the star with the rotating
Geneva isochrones, we get an age of 10Myr (pre blue-loop) or 8Myr
(post blue-loop). For the specific models used in this analysis, the
masses of RSGs in a 12Myr-old cluster do not experience a
blue-loop. However, the specific masses of stars which experience
blue-loops are extremely sensitive to the input physics, such as
rotation \citep[see Fig.~1 of][]{Hirschi04}. Hence, a blue-loop may be
introduced for stellar initial masses relevant to RSGC1 simply by
changing the rotational speed. Additionally, it is likely that
blue-loops would be affected by the inclusion of extra physics (e.g.\
magnetic fields). In summary, We do not necessarily interpret F15's
location in the H-R diagram as evidence of cluster non-coevality, it
could simply be that the input physics of the evolutionary models used
in the analysis are not fine-tuned to this cluster.

If the RSGs of this cluster do experience some form of blue-loop, then
the position of F15 on the H-R diagram is consistent with it evolving
{\it away} from the RSGs. A post-RSG nature for this star would make
it a member of a very exclusive club -- arguably IRC +10420 is the
only object which is widely accepted to be a post-RSG, though a case
has also been argued for HD~179821 \citep[see review
of][]{Oudmaijer08}. We note that F15 does not exhibit the same
considerable IR excess, nor the bright maser emission of IRC
+10420. This may be due to F15's lower initial mass -- the Geneva
models imply $\sim$18\msun\ for F15, while the larger luminosity of IRC
+10420 makes it consistent with a star of initial mass $\sim$40\msun\
(see also Sect.\ \ref{sec:masers}).

The cluster age we derive is slightly greater than the age of
$\la$9Myr derived in FMR06. This previous estimate was determined by
comparing the luminosity spread of the RSGs to that predicted by the
non-rotating isochones of \citet{Schaller92} as a function of age. The
more rigourous investigation of the stars' luminosities in this present
paper results in a larger luminosity spread for the RSGs, while the
non-rotating models do not reproduce the higher luminosities. Much
better agreement is found between the contemporary rotating models and
the new luminosity estimates. 

Finally, we remark that the observed temperatures of the stars
are systematically cooler than the isochrones (see also Fig.\
\ref{fig:masers}). This could be reconciled by increasing either the
relative metal abundances or the stellar rotational velocities. Both
lead to slightly increased stellar radii, the former due to the
increased opacity of the envelope, the latter due to the lower
effective gravities. A super-solar metallicity would certainly be
consistent with the Galactic metallicity gradient and the cluster's
Galacto-centric distance ($\sim$3\,kpc). However, given the well-known
disparities between observations and theory in the field of RSGs, we
attach a cautionary note to any conclusions derived from this
evidence. While recent progress has been made in uniting theory and
observation at solar metallicity \citep{Levesque05}, discrepancies
still exist at sub-solar metallicities \citep{Levesque06}. The
location in the Galaxy of the Scutum clusters would seem to make
non-solar metallicities likely. Accurate abundance measurements of the
clusters would make them ideal testbeds for evolutionary models, as
well as probes of the Galactic metallicity gradient.

\subsection{Cluster mass} \label{sec:mass}
To determine the cluster mass we employ the same Monte-Carlo technique
used in FMR06 and DFK07. We generate a synthetic cluster of a
pre-defined initial mass, containing stars whose masses are randomly
drawn from a distribution consistent with a Salpeter initial mass
function \citep{Salpeter55}. Then, for a given cluster age, we
determine the present-day masses, temperatures and luminosities from
the Geneva isochrones used in Sect.\ \ref{sec:age}. We then count the
number of RSGs in the cluster, where we define a RSG as a star whose
temperature is lower than 4000K and luminosity greater than
$10^{4}$\lsun. As this is a random process, we repeat each simulation
1000 times to reduce statistical noise.

\begin{figure}[t]
  \centering
  \includegraphics[width=12cm,bb=20 10 570 450]{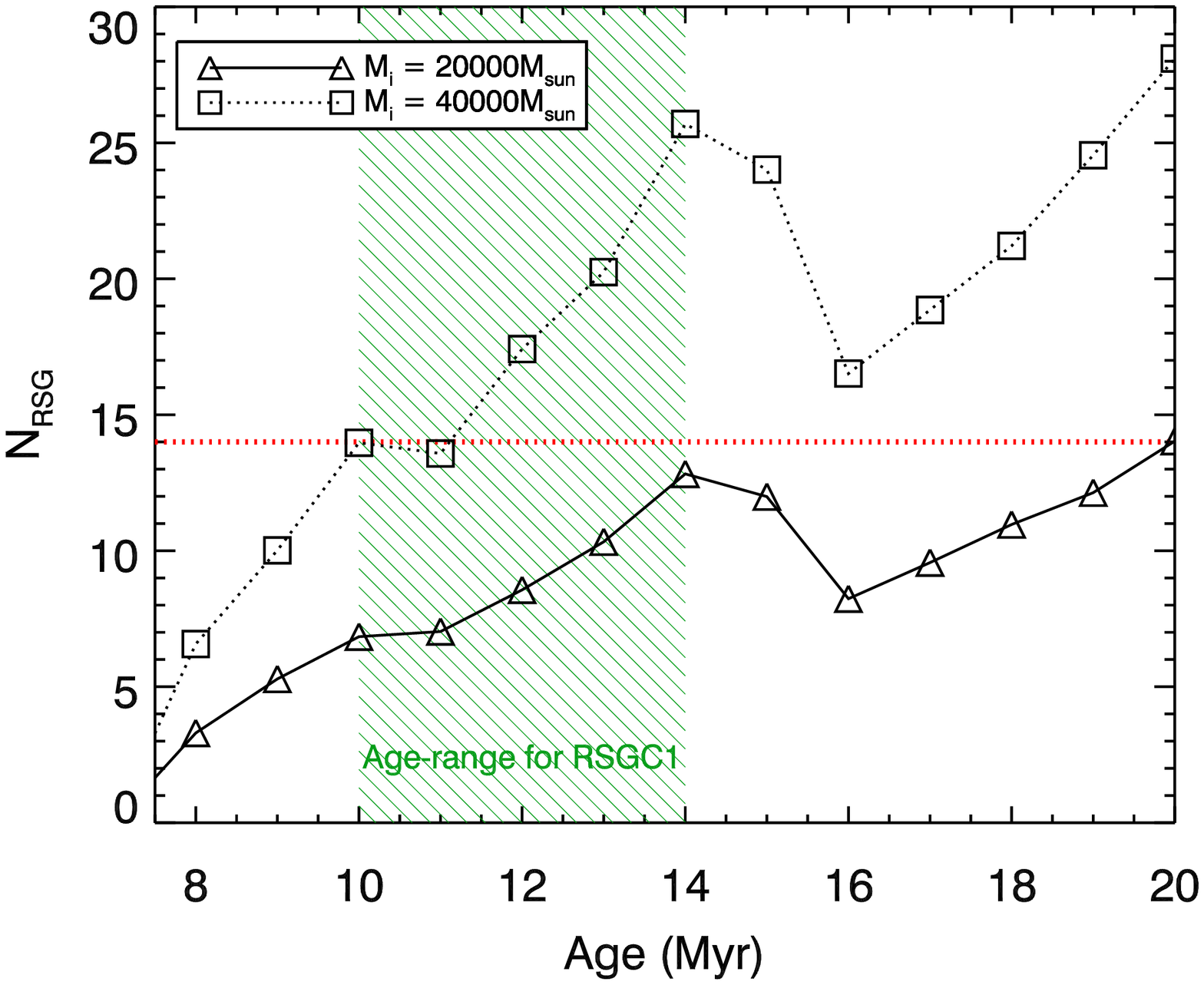}
  \caption{The number of RSGs as a function of age for a coeval
  cluster of a given initial mass, calculated using the rotating
  Geneva isochrones \citep{Mey-Mae00}. The age-range for RSGC1,
  determined from isochrone fitting, is indicated on the plot, as is
  the number of RSGs observed in the cluster.}
  \label{fig:mass_evo}
\end{figure}

In Fig.\ \ref{fig:mass_evo}, we plot the number of RSGs contained in a
synthetic cluster as a function of cluster age, for two initial
cluster masses: 20,000\msun\ and 40,000\msun. The plot shows that
significant numbers of RSGs begin to be seen after $\sim$7Myr -- in
clusters younger than this the post-MS stars are massive enough to
evolve directly to the WN phase, skipping the RSG stage. Peaks in the
number of RSGs are reached at $\sim$14Myr and $\sim$20Myr, the dip in
between is caused by the onset of a blue-loop in the stars' evolution
for a narrow range of initial masses. The mean luminosity of the RSGs
as a function of time decreases, as the initial masses of the stars in
the RSG zone becomes smaller (see Fig.\ \ref{fig:hrd}, and discussion
in Sect.\ \ref{sec:clusters}). For a cluster containing 14 RSGs with
an inferred age of 12$\pm$2\,Myr (see above), we find that the initial
mass of the cluster must be somewhere in between these two, implying
an initial mass of RSGC1 of $(3 \pm 1) \times 10^{4}$\msun.

Using the velocity dispersion of the RSGs, we can compare this value
of the cluster's initial mass to the cluster's {\it dynamical} mass,
under the assumption that the cluster is currently in virial
equilibrium. The dynamical mass $M_{\rm dyn}$ is derived using the
relation,

\begin{equation}
M_{\rm dyn} = \frac{\eta \sigma^{2}_{v} r_{\rm hp}}{G}
\label{equ:dynmass}
\end{equation}

\noindent where $r_{\rm hp}$ is the half-light radius,
$\sigma^{2}_{v}$ is the velocity dispersion, $G$ is the gravitational
constant, and $\eta$ is a constant which depends on the stellar
distribution with radius, and is typically taken to be $\sim$10
\citep[see review in Introduction of][]{Mengel02}. 

\begin{figure}[t]
  \centering
  \includegraphics[width=12cm,bb=50 15 700 530]{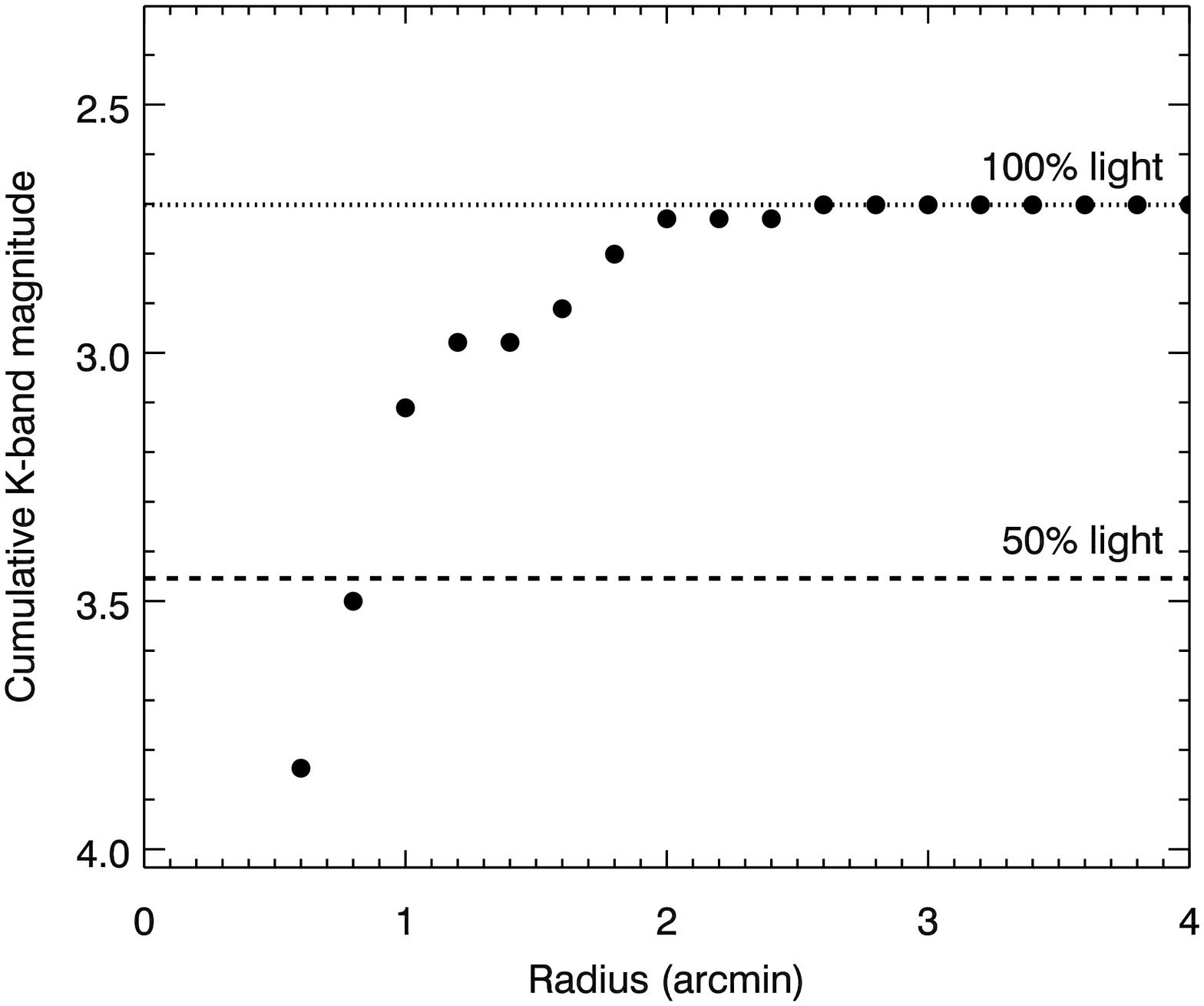}
  \caption{Cumulative apparent magnitude distribution of the cluster,
  centred on 18$^{h}$37$^{m}$57.4$^{s}$, -6\degr52\arcmin58.11\arcsec\
  (J2000), using the 15 cool luminous stars as tracers of the
  cluster's spatial distribution.  }
  \label{fig:halflight}
\end{figure}

For the velocity dispersion, we find $\sigma^{2}_{v} = 3.7$\kms\ after
the internal uncertainty in the wavelength solution ($\pm$1\kms) has
been subtracted in quadrature. To find the half-light radius, we plot
the cumulative brightness profile of the cluster using stars F01--15
as tracers of the cluster's spatial distribution (see Fig.\
\ref{fig:halflight}). We used a cluster centre of
18$^{h}$37$^{m}$57.4$^{s}$, -6\degr52\arcmin58.11\arcsec\ (J2000), the
approximate mid-point of the cool stars. We experimented with moving
the cluster centre by up to 0.5\arcmin, using different photometric
bands, and using the luminosities derived in Sect.\ \ref{sec:lum}
rather than the raw photometry. From all these methods we found that
the cluster half-light radius was stable at $0.8 \pm 0.1$\arcsec. At
the distance derived in Sect.\ \ref{sec:dist}, this gives a cluster
size $r_{\rm hp} = 1.5 \pm 0.3$pc.

Using these values, we find the dynamical mass of RSGC1 to be {$M_{\rm
dyn} = (5 \pm 1) ~\eta/10 \times 10^{4}$\msun.} Due to the extra
uncertainty in the density parameter $\eta$, we consider this to be an
order-of-magnitude estimate only, which compares well to the initial
cluster mass derived using evolutionary models.

\begin{table}[t]
  \begin{center}
    \caption{Properties of the two Scutum RSG clusters. (1): cluster
    name; (2): total initial cluster mass derived from evolutionary
    models; (3): cluster mass derived from stellar velocity
    dispersion; (4): cluster age; (5): cluster diameter; (6): cluster
    distance; (7): distance from cluster to the Galactic centre; (8):
    initial mass range of the RSGs within the clusters. RSGC2 values
    come from DFK07.}
    \begin{tabular}{l|cc|c|c|cc|c}
\hline
(1) & (2) & (3) & (4) & (5) & (6) & (7) & (8) \\
Cluster & $M_{\rm init}$ (evol) & $M_{\rm dyn}$ & Age & $d$  & $D_{\odot}$
& $D_{\rm GC}$ & $M_{\rm init} (RSGs) $\\
        & \multicolumn{2}{c|}{($\times10^{4}$\msun)} & (Myr) & (pc) &
\multicolumn{2}{c|}{(kpc)} & (\msun) \\
\hline \hline
RSGC1   & 3$\pm$1 & 5$\pm$1 & 12$\pm$2 & 1.5$\pm$0.3 & 6.60$\pm$0.89 & 3.2 & 18$^{+4}_{-2}$ \\
RSGC2   & 4$\pm$1 & 6$\pm$4 & 17$\pm$3 & 3.2$^{+1.2}_{-0.7}$ & 5.83$^{+1.91}_{-0.76}$ & 3.5 & 14$\pm$2 \\
\hline
    \end{tabular}
    \label{tab:clusters}
  \end{center}
\end{table}

\subsection{Comparison of the two RSG clusters} \label{sec:clusters}
Given that, until recently, the largest number of RSGs in any one
cluster was 5, the discovery of the two Scutum RSG clusters lying so
close to one another is remarkable. After applying the same analysis
techniques to each cluster, we summarize the physical parameters of
RSGC1 and RSGC2 in Table \ref{tab:clusters}. 

The radial velocities of each cluster put them at the tangential point
of the Galactic rotation curve, with Galacto-centric distances of
3-4kpc. The clusters are separated from one another by
$0.8^{+1.6}_{-0.7}$\,kpc. Their proximity, combined with their similar
ages, suggests that they were both formed in a region-wide starburst
phase some 10-20Myr ago, and that the chemical abundances of their
natal material should be similar. 

First-order evidence of uniform metallicity between the clusters comes
from the median spectral-types of the RSGs, which is M3 in each
cluster. The average spectral-type of RSGs has been shown to be
dependent on environment, shifting gradually to later types with
increasing metallicity -- averages of K5, M1, and M2 were found for
the SMC, LMC and Galaxy respectively \citep{Elias85,M-O03}. Proposed
physical explanations for this include (a) reduced metallicity leading
to lower envelope opacities, increased stellar radii and hence to
systematically lower effective temperatures; or (b) the effect of
lower metal abundances on the strengths of the diagnostic TiO
lines. Regardless, the average spectral-type of M3 for the two Scutum
clusters would seem to suggest that they have similar, possibly
super-solar metallicities. The chemical abundances of the stars in
these clusters will be the subject of a future paper.

Both from the analysis of evolutionary models, and the velocity
dispersion, we find similar masses for each cluster -- the factor of
two difference in the number of RSGs is caused by the difference in
cluster ages. This may also explain the slightly larger size of RSGC2:
it has been suggested by \citet{B-G06} that a cluster of this initial
mass and age may be {\it out} of virial equilibrium. This arises when
the left-over natal material is expelled from the cluster by the first
SNe explosions of the cluster's most massive stars. This leaves the
remaining stars with a velocity dispersion larger than that of a
virialized cluster, resulting in cluster expansion. This may explain
(i) the slightly different sizes of the clusters, (ii) the apparent
lack of any obvious diffuse nebular emission in the GLIMPSE
images. Also, we note that the dynamical masses of each cluster are
slightly higher than the `evolutionary' masses, though the errors on
the dynamical masses, in particular the density parameter $\eta$, make
it difficult to draw any firm conclusions from this.

\begin{figure}[t]
  \centering
  \includegraphics[width=12cm,bb=20 20 542 538]{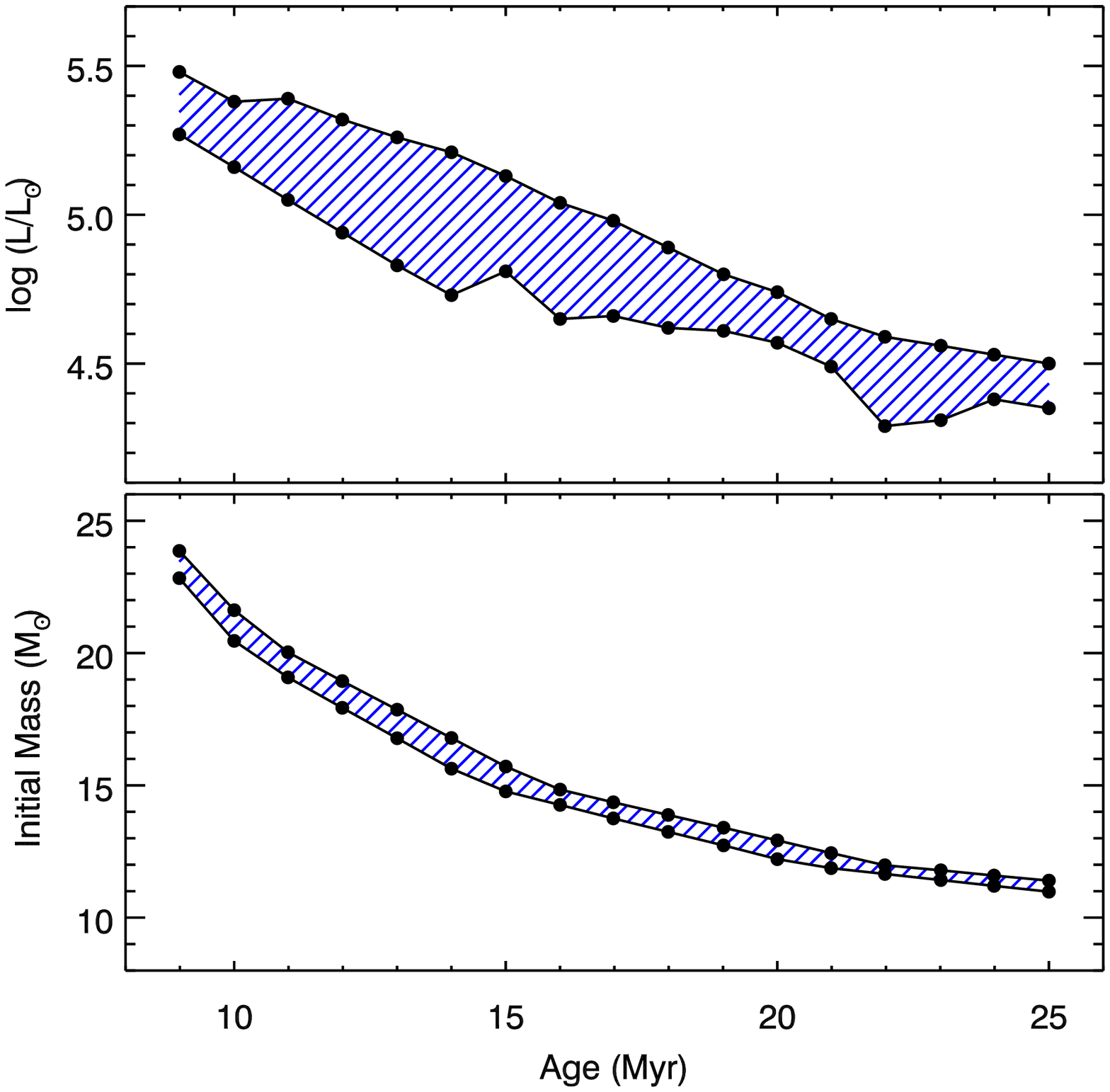}
  \caption{The minimum and maximum luminosities (top panel) and
  initial masses (bottom panel) of stars with effective temperatures
  cooler than 4000K in a coeval cluster. Calculated using the rotating
  Geneva isochrones of \citet{Mey-Mae00}.  }
  \label{fig:minmaxmass}
\end{figure}

The difference in ages between the two clusters implies that the
initial masses of the RSGs in each must be different; in the younger
RSGC1 we are seeing stars with larger initial masses than in the
slightly older RSGC2.  To investigate the initial masses of the stars
in each cluster, we again use the evolutionary models of
\citet{Mey-Mae00}. For a given isochrone, we find the minimum and
maximum initial masses and luminosities of all stars with effective
temperatures cooler than 4000K, i.e.\ the temperature range of
the RSGs. In Fig.\ \ref{fig:minmaxmass} we plot the luminosity and
mass ranges of cool stars as a function of time. From the luminosity
ranges of the stars in each cluster, we then determine the initial
masses of the RSGs. For RSGC1 we find $M_{\star, \rm init} =
18^{+4}_{-2}$\msun, and for RSGC2 $M_{\star, \rm init} = 14 \pm
2$\msun. This underlines the potential importance of the RSG clusters
to the study of stellar evolution: {\it they allow us to study large
numbers of RSGs at uniform metallicity as a function of initial mass.}

\subsection{Masers in RSGs} \label{sec:masers}
One aspect of RSG evolution which the two Scutum RSG clusters allow us
to study is the onset of the maser-active phase. Masers are often
observed in `extreme' cool stars, e.g.\ the RSGs VY~CMa,
NML~Cyg, S~Per \citep[see][ and refs therein]{R-Y98}, which are also
synonymous with large IR excesses. In the standard picture, these
phenomena are caused by episodes of high mass-loss, producing large
amounts of circumstellar material which gives rise to the IR
excess. The masers themselves originate in the outflowing
material. The SiO 43GHz maser is formed low in the wind at higher
temperatures; at larger radii the formation of dust grains leads to a
depletion of SiO. In the SiO maser-forming region the outflow velocity
is low, and hence SiO masers typically have radial velocities similar
to the stellar systemic velocity, $v_{\rm sys}$
\citep{Jewell91}. Higher in the wind the H$_2$O 22GHz maser forms, and
typically has peaks at many velocities between $v_{\rm sys} \pm
v_{\infty}$, where $v_{\infty}$ is the wind's terminal velocity
(typically around 10-30\kms). At larger radii still, H$_2$O is
photo-disassociated to OH, giving rise to the 1612MHz OH maser. Here
the outflow has reached its terminal velocity, giving the line-profile
its typical double-peaked morphology (where the separation of the
peaks is 2$v_{\infty}$ and the centroid is $v_{\rm sys}$). For a more
comprehensive review of masers in luminous cool stars, see
\citet{Habing96}.

\begin{figure}[t]
  \centering
  \includegraphics[width=12cm,bb=10 10 520 500,clip]{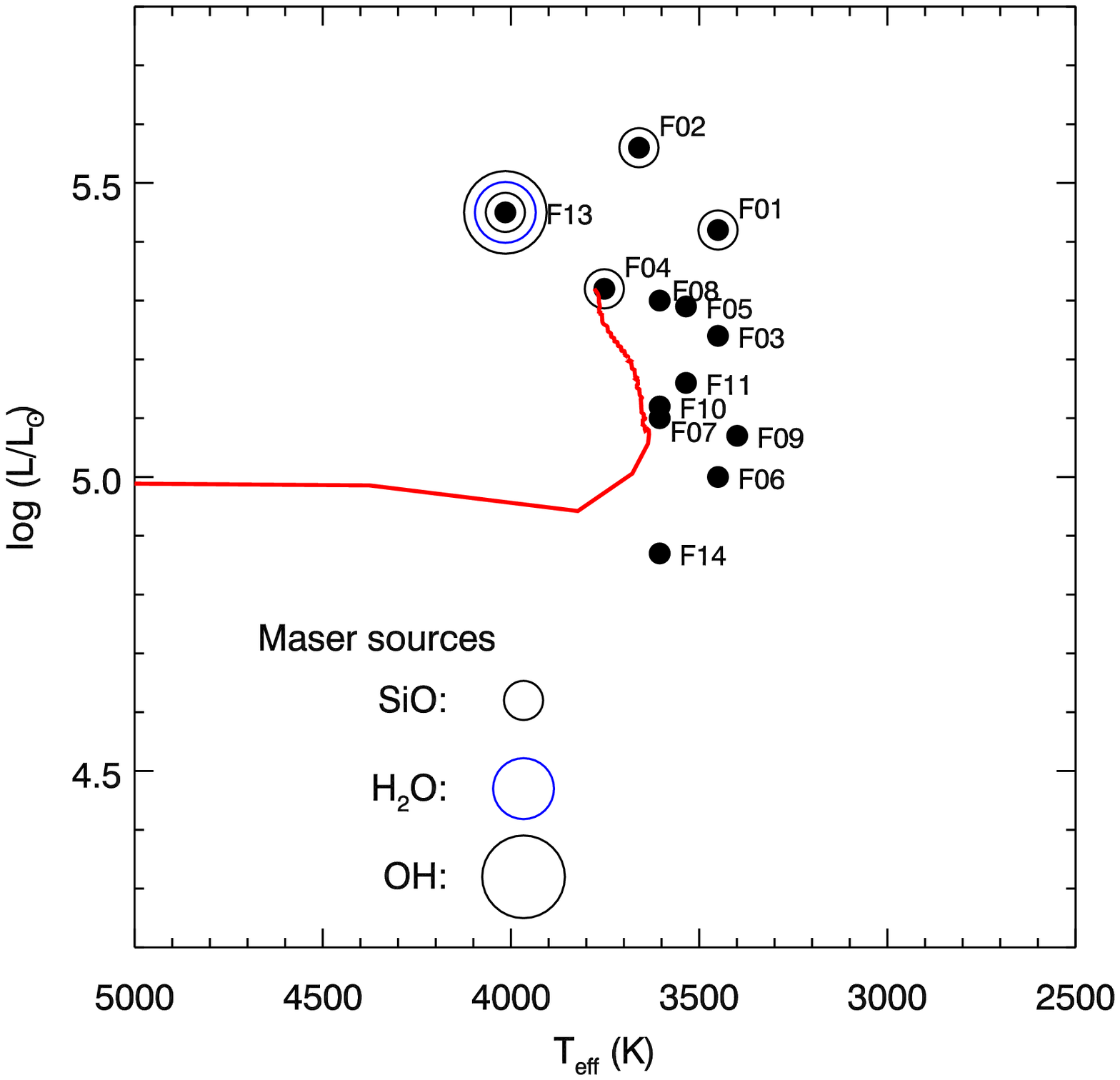}
  \caption{H-R diagram of the RSGs in RSGC1, showing those stars which
  are maser sources. A 12Myr model isochone is overplotted, from the
  rotating Geneva models of \citet{Mey-Mae00}.}
  \label{fig:masers}
\end{figure}

It is unclear as to whether the maser stage is one which all RSGs will
go through, or whether only extreme objects will pass through this
phase. Much work has been done on the detailed physical conditions
under which masers form \citep[see review of][]{Habing96}, however in
a simplified picture one may say that the presence of the different
masers is determined by the wind density, i.e. by mass-loss rate and
by pulsations. For an increasing mass-loss rate, gradually the
critical density will be reached in the formation zones of each
transition; while stellar pulsations will create density-contrasts in
the outflow, conducive to population inversions. That is to say that
we expect to see maser emission from those RSGs which have the
strongest mass-loss rates and which are pulsationally unstable. 

Mass-loss from RSGs is driven by radiation pressure on dust grains in
the outer atmosphere, and hence depends upon the star's effective
temperature (cool enough to allow dust to form) and luminosity (high
radiation pressure) \citep{v-L05}. Hence we may expect to find masers
in high-mass RSGs which have the higher luminosities, while among a
coeval sample of RSGs one may expect to see masers in those objects
furthest along their evolution, where their path on the HR diagram
takes a sharp upturn (see Fig.\ \ref{fig:hrd}). Further, as
masers are linked to stellar pulsations, we may expect to find masers
in unstable stars -- i.e.\ stars with high $L_{\star}/M_{\star}$
ratios that are evolving closer to the Humphreys-Davidson limit at
$L_{\star} \sim 10^{5.7}$\lsun\ \citep{H-D79}. This zone of the H-R
diagram is often linked to the so-called `modified' Eddington limit,
when the contributions of atomic/molecular transitions to the
continuum opacity are included when calculating the Eddington
luminosity.

To investigate the presence of masers in RSGs, in Fig.\
\ref{fig:masers} we plot a HR diagram of the RSGs in RSGC1, and
indicate those stars which are maser sources. In addition to the OH
maser observations presented here, SiO and H$_2$O maser observations
of RSGC1 were also taken by \citet{N-D06}. They found that stars F01,
F02, F04 and F13 were spatially coincident with SiO emission, while
they concluded that the H$_{2}$O maser emission they found was likely
to come from F13\footnote{The large beamsize of the 22GHz H$_{2}$O
maser, 73\arcsec, overlaps several other stars in the
cluster. However, as these masers are often observed in stars with
maser emission from SiO, and as the central velocity of the H$_{2}$O
maser was consistent with that from the SiO maser emission of F13,
\citet{N-D06} concluded that F13 was likely to be the origin of the
H$_{2}$O maser emission.  }.

The figure shows that it is the most luminous stars of the cluster
which exhibit maser emission. Further, F13, which is the source of
SiO, H$_{2}$O and OH masers, appears to be at the point of evolving
back toward the blue. The figure therefore appears to support the
hypothesis that maser emission activates (or becomes strong enough to
be observed) in the latest stages of RSG evolution, when the star's
mass-loss rate is highest and the star becomes unstable to
pulsations. 

This hypothesis could be tested with a study of the second Scutum RSG
cluster, RSGC2. This cluster also has many RSGs and so we are again
seeing stars in both the earlier and later RSG stages. As mentioned in
Sect.\ \ref{sec:clusters}, the stars in RSGC2 are less luminous and
hence further from the H-D limit. Though they likely have lower
initial masses, they have lower $L_{\star}/M_{\star}$ ratios. Thus, we
may not expect to see maser emission from these stars, especially
those with lower $L_{\star}$. A comprehensive maser study of the RSGC2
region would be able to test at which phase of RSG evolution stars
become maser-active, while non-detections would place lower-limits to
the initial-mass requirements to pass through the maser-active phase.

Finally, we note that the YHG F15 is not observed to have maser
emission, unlike the prototype post-RSG IRC +10420. Masers are rarely
observed around stars hotter than $\sim$4000, and the presence of
maser emission in IRC +10420's outflow is commonly accepted as
evidence of its rapid evolution away from the RSG phase. That no maser
is observed in F15 may be indicative of a lower wind-density while in
the RSG phase, due to its lower initial mass ($\sim$18\msun\ compared
to $\sim$40\msun\ for IRC +10420).


\section{Conclusions}
We have presented a comprehensive investigation into the physical
properties of the luminous cool stars in the massive cluster
RSGC1. Using high-resolution spectroscopy we have accurately measured
the cluster's radial velocity, derived its kinematic distance, and
reappraised the stars' temperatures and luminosities. We find a larger
luminosity spread than in the discovery paper, which is well fitted by
a cluster age of 12Myr and cluster mass of $(3 \pm 1) \times
10^4$\msun. The mass is similar to that of the nearby cluster RSGC2,
and we suggest that the difference in the number of RSGs in each is
due to the separation in cluster ages, with RSGC2 being somewhat
older. This implies that the initial masses of the RSGs in each
cluster are different, which we determine to be $\sim$18\msun\ for
RSGC1 and $\sim$14\msun\ for RSGC2. The clusters therefore allow the
study of RSG evolution as a function of initial mass, while
constraining the variable of metallicity. Finally, with new 1612MHz
radio observations we find compelling evidence that the OH maser is
associated with star F13, and collating recent maser observations of
the cluster we argue that the maser-active phase is associated with
stars in the latter stages of RSG evolution.

\acknowledgments Acknowledgments: We would like to thank Maria
Messineo for useful discussions concerning masers in cool stars, and
the anonymous referee for comments and suggestions which improved the
paper. We thank Mark Claussen for discussions regarding the
astrometric accuracy of the VLA observations. The material in this
work is supported by NASA under award NNG~05-GC37G, through the
Long-Term Space Astrophysics program. This research was performed in
the Rochester Imaging Detector Laboratory with support from a NYSTAR
Faculty Development Program grant.  The National Radio Astronomy
Observatory is a facility of the National Science Foundation operated
under cooperative agreement by Associated Universities, Inc.  Part of
the data presented here were obtained at the W.\ M.\ Keck Observatory,
which is operated as a scientific partnership among the California
Institute of Technology, the University of California, and the
National Aeronautics and Space Administration. The Observatory was
made possible by the generous financial support of the W.\ M.\ Keck
Foundation. This publication makes use of data products from the Two
Micron All Sky Survey, which is a joint project of the University of
Massachusetts and the Infrared Processing and Analysis
Center/California Institute of Technology, funded by the National
Aeronautics and Space Administration and the NSF. This research has
made use of MSX and Spitzer's GLIMPSE survey data, the {\sc simbad}
database, Aladin \& IDL software packages, and the GSFC IDL library.

\clearpage
\bibliographystyle{/fat/Data/model_paper/aa}
\bibliography{/fat/Data/bibtex/biblio}

\end{document}